\documentclass[aps,prb,twocolumn,groupedaddress]{revtex4}

\usepackage{graphicx}
\usepackage{dcolumn}
\usepackage{bm}
\usepackage{amssymb}

\begin{document}

%\preprint{}

%Title of paper
\title{ 
Structure formation in binary mixtures of surfactants: vesicle opening-up to bicelles and octopus-like micelles}

\author{Hiroshi Noguchi}
\email[]{noguchi@issp.u-tokyo.ac.jp}
\affiliation{
Institute for Solid State Physics, University of Tokyo,
 Kashiwa, Chiba 277-8581, Japan}
%\homepage[]{Your web page}
%\altaffiliation{}

\date{\today}

\begin{abstract}
Micelle formation in binary mixtures of surfactants is studied 
using a coarse-grained molecular simulation.
When a vesicle composed 
of lipid and detergent types of molecules is ruptured, 
a disk-shaped micelle, the bicelle, is typically formed.
It is found that cup-shaped vesicles and
bicelles connected with worm-like micelles
are also formed depending on the surfactant ratio and 
critical micelle concentration.
The obtained octopus shape of micelles agree  with those observed in the cryo-TEM images reported
in  [S. Jain and F. S. Bates, Macromol. 37, 1511 (2004).].
Two types of connection structures between the worm-like micelles and the bicelles 
are revealed.
\end{abstract}

%\pacs{87.16.Dg Membranes, bilayers, and vesicles 
%      87.17.Aa Theory and modeling; computer simulation
%      82.70.Uv Surfactants, micellar solutions, vesicles, lamellae, 
%               amphiphilic systems
%     }
%\pacs{87.16.Dg, 87.17.Aa, 82.70.Uv}
%\keywords{}

\maketitle

\section{Introduction}

Amphiphilic molecules, such as lipids and detergents,
self-assemble various structures in aqueous solutions \cite{isra11,safr94,jain03,sedd04,caff09,ujwa11,walt91,yama09,jain04}.
In solutions of a single type of simple surfactant,
the structure of the aggregates is qualitatively understood
by the relative
size of their hydrophilic parts \cite{isra11,safr94}.
Cone-shaped molecules form spherical or worm-like micelles,
cylindrical molecules  form bilayer membranes, and
inverted cone-shaped molecules  form
inverted hexagonal structures, or inverted micelles.
Mixtures of different types of surfactants
show more variety of self-assembled structures than single-component systems.
The edges of bilayer membranes can be stabilized by cone-shaped surfactants.
Hence, discoidal bilayer membranes called bicelles (bilayered micelles) 
or pores on the membranes are formed \cite{sedd04,caff09,ujwa11,walt91,yama09}.
Recently, an undulating cylinder and
a bilayer disk connected with several worm-like micelles, which shape looks like an octopus or jelly-fish,
have been observed in binary mixtures of diblock copolymers by cryo-TEM \cite{jain04}.
When compared with the structures of single-component systems,
multi-component self-assembled structures are not well understood.

Mixtures of lipids and detergents are used for solubilization, reconstruction, and crystallization of membrane proteins \cite{sedd04,caff09}.
In particular, a bicelle  is an excellent medium to investigate protein functions \cite{sand98,ujwa11}. 
The proteins maintain their functionality in the bicelles \cite{czer00},
and magnetically aligned bicelles are used for NMR studies \cite{ange07}.
The structures of several proteins have been determined by the bicelle method \cite{ujwa11,faha02}.

The dynamics of  micelle-to-vesicle transitions
 have been studied experimentally by time-resolved scatting techniques (light, neutron, 
and X-ray) \cite{leng02,weis05,brys05,made11,gumm11}.
Typically, the surfactants aggregate into disk-like 
micelles, which grow and
transform into vesicles when their radii exceed a critical size \cite{from83}.
Since fusion between vesicles is very slow, the size of the first-formed vesicles 
is maintained in the time scale of experiments.
Thus, the vesicle size is determined by kinetics rather than by thermal equilibrium.
In the opposite transition from vesicles to micelles, 
various pathways of the solubilization of liposomes are observed depending on 
the types of lipids and detergents \cite{nomu01,sudb11,tomi11,elsa11,hama12}:
Rhythmic shrinkage, bursting, budding, peeling, and inside-out inversion.

Recently, multi-component bilayer membranes
have been intensively investigated, since they are believed to play important roles in living cells.
Several patterns of phase-separated domains have been observed on giant liposomes \cite{baum03,veat03,gudh07,coll08,yana08,chri09,baga09},
and the spontaneous curvature produced by proteins or glycolipids
induce a local high curvature such as in a membrane tube \cite{baum10,phil09,shny09,akiy03}.
The domain formations and morphological changes of the vesicles can be observed by optical microscopy.
However, the structure dynamics of micelle formation are much more difficult
to investigate; since their size are  $10$nm scale, electron microscopy has to be used for direct observation.

When molecular simulations are used, the structure formation in molecular scales can be revealed in detail.
In contrast to the fact that many simulations have been performed on domain formation in membranes\cite{baga09,tani96,kohy03,lara05,huan11,hu11},
the structure formation of micelles is much less explored.
In this paper, we study various structures of binary mixtures using a coarse-grained molecular simulation.

Although computer technology has grown,
the typical scale for recent simulations of atomistic models
is only $100$ ns dynamics of hundreds of lipid molecules.
In order to simulate bilayer membranes on longer and larger scales,
various types of coarse-grained molecular models have been proposed (see review articles \cite{muel06,vent06,nogu09,marr09}).
Recently, the bottom-up approaches were taken to construct coarse-grained molecular models, 
where potential parameters are tuned from atomistic simulations \cite{marr04,izve05,arkh08,shin08,wang10,shin11}.
On the other hand, we choose a top-down approach to construct coarse-grained molecular models, 
where potentials are based on the continuum theory.
Here,  one of the solvent-free molecular models \cite{nogu11} is employed.
In this model, the membrane properties such as the bending rigidity and spontaneous curvature of the monolayer
 can be varied over wide ranges.

In Sec. \ref{sec:method}, the surfactant model and simulation method are described.
In Sec. \ref{sec:plane}, the membrane properties of the phase boundary are investigated
using the planar membranes with straight edges or pores.
The line tension of the phase boundary is estimated in Sec. \ref{sec:line_ab}.
The phase segregation around the membrane edges
and the pore shapes on the phase boundary are described in  Secs. \ref{sec:wet} and \ref{sec:edge}, respectively.
In  Sec. \ref{sec:openup}, the opening-up of the two-component vesicles are explored.
Various structures including octopus-like micelles are obtained.
The summary and discussion are given in Sec. \ref{sec:sum}.

\section{Simulation Model and Method}
\label{sec:method}

\subsection{Surfactant Model}

A solvent-free molecular model \cite{nogu11} is used to simulate two-component surfactant mixtures.
We consider  $N_{\rm A}$ molecules of type A and  $N_{\rm B}$ molecules of type B.
The type B molecule is a lipid with a cylindrical shape (the spontaneous curvature of the monolayer $C_0 =0$).
The type A molecule is a model of a detergent or lipid molecule with a cylindrical or cone shape  with a relatively
large head ($C_0 \ge 0$).

Each ($i$-th) molecule has a spherical particle with an orientation vector ${\bf u}_i$, 
which represents the direction from the hydrophobic to the hydrophilic part.
There are two points of interaction in the molecule:
the center of a sphere ${\bf r}^{\rm s}_i$ and a hydrophilic point ${\bf r}^{\rm e}_i={\bf r}^{\rm s}_i+{\bf u}_i \sigma$.
The molecules interact with each other via the potential,
\begin{eqnarray}
\frac{U}{k_{\rm B}T} &=\ \ & \hspace{1cm} \sum_{i<j} U_{\rm {rep}}(r_{ij}^{\rm s}) \label{eq:U_all}
               + \sum_{i} \varepsilon_i\  U_{\rm {att}}(\rho_i) \nonumber \\ \nonumber
&\ \ +& \ \ \frac{k_{\rm{tilt}}}{2} \sum_{i<j} \bigg[ 
( {\bf u}_{i}\cdot \hat{\bf r}^{\rm s}_{ij})^2
 + ({\bf u}_{j}\cdot \hat{\bf r}^{\rm s}_{ij})^2  \bigg] w_{\rm {cv}}(r^{\rm e}_{ij}) \\ \nonumber
&\ \ +&  \frac{k_{\rm {bend}}}{2} \sum_{i<j}  \bigg({\bf u}_{i} - {\bf u}_{j} - C_{\rm {bd}}^{ij} \hat{\bf r}^{\rm s}_{ij} \bigg)^2 w_{\rm {cv}}(r^{\rm e}_{ij}) \\ 
&\ \ +&   \varepsilon_{\rm {AB}} \sum_{i<N_{\rm A}, j \ge N_{\rm A}} U_{\rm {AB}}(r_{ij}^{\rm s}),
\end{eqnarray} 
where ${\bf r}_{ij}={\bf r}_{i}-{\bf r}_j$, $r_{ij}=|{\bf r}_{ij}|$,
 $\hat{\bf r}_{ij}={\bf r}_{ij}/r_{ij}$, and $k_{\rm B}T$ is the thermal energy.
Here, type A molecules are considered for $0 \le i< N_{\rm A}$ and type B for  $N_{\rm A} \le i< N=N_{\rm A}+N_{\rm B}$.

The molecules have an excluded volume with a diameter $\sigma$ via the repulsive potential,
$U_{\rm {rep}}(r)=\exp[-20(r/\sigma-1)]$,
with a cutoff at $r=2.4\sigma$.
The second term in Eq. (\ref{eq:U_all}) represents the attractive interaction between the molecules.
The two types of molecules can have different attractive strengths:
$\varepsilon_i = \varepsilon_{\rm att}^{\rm A}$  for $0 \le i< N_{\rm A}$ 
and $\varepsilon_i = \varepsilon_{\rm att}^{\rm B}$  for $N_{\rm A} \le i< N$.
An attractive multibody potential $U_{\rm {att}}(\rho_i)$ is 
employed to mimic the ``hydrophobic'' interaction.
The potential $U_{\rm {att}}(\rho_i)$ is given by
\begin{equation} \label{eq:U_att}
U_{\rm {att}}(\rho_i) = 0.25\ln[1+\exp\{-4(\rho_i-\rho^*)\}]- C,
\end{equation}
with $C= 0.25\ln\{1+\exp(4\rho^*)\}$.
The local particle density $\rho_i$ is approximately the number of
particles ${\bf r}^{\rm s}_i$  in the sphere with radius $r_{\rm {att}}$.
\begin{equation}
\rho_i= \sum_{j \ne i} f_{\rm {cut}}(r^{\rm s}_{ij}), 
\label{eq:wrho}
\end{equation} 
where $f_{\rm {cut}}(r)$ is a $C^{\infty}$ cutoff function~\cite{nogu06},
\begin{equation} \label{eq:cutoff}
f_{\rm {cut}}(r)=\left\{ 
\begin{array}{ll}
\exp\{A_0(1+\frac{1}{(r/r_{\rm {cut}})^n -1})\}
& (r < r_{\rm {cut}}) \\
0  & (r \ge r_{\rm {cut}}) 
\end{array}
\right.
\end{equation}
with $n=6$, $A_0=\ln(2) \{(r_{\rm {cut}}/r_{\rm {att}})^n-1\}$,
$r_{\rm {att}}= 1.9\sigma$  $(f_{\rm {cut}}(r_{\rm {att}})=0.5)$, 
and the cutoff radius $r_{\rm {cut}}=2.4\sigma$.
The density $\rho^*$ in $U_{\rm {att}}(\rho_i)$ is the characteristic density,
above which 
the pairwise attractive interaction is smoothly cut off.

The third and fourth terms in Eq.~(\ref{eq:U_all}) are
discretized versions of the 
tilt and bending potentials of the tilt model \cite{hamm98,hamm00}, respectively.
This type of bending potential was
also used in other coarse-grained lipid models \cite{nogu03,fara09}.
A smoothly truncated Gaussian function~\cite{nogu06} 
is employed as the weight function 
\begin{equation} \label{eq:wcv}
w_{\rm {cv}}(r)=\left\{ 
\begin{array}{ll}
\exp (\frac{(r/r_{\rm {ga}})^2}{(r/r_{\rm {cc}})^n -1})
& (r < r_{\rm {cc}}) \\
0  & (r \ge r_{\rm {cc}}) 
\end{array}
\right.
\end{equation}
with  $n=4$, $r_{\rm {ga}}=1.5\sigma$, and $r_{\rm {cc}}=3\sigma$.
The spontaneous curvatures of the monolayer membranes
are varied by parameters $C_{\rm {bd}}^{\rm A}$ and $C_{\rm {bd}}^{\rm B}$:
$C_{\rm {bd}}^{ij} =  (C_{\rm {bd}}^{i}+ C_{\rm {bd}}^{j})/2$,
 $C_{\rm {bd}}^{i} = C_{\rm {bd}}^{\rm A}$  for $0 \le i< N_{\rm A}$, 
and $C_{\rm {bd}}^{i} = C_{\rm {bd}}^{\rm B}$  for $N_{\rm A} \le i< N$.
A single-component monolayer membrane has spontaneous curvature 
$C_0  \simeq \{k_{\rm {bend}}/(k_{\rm {bend}}+k_{\rm {tilt}})\}C_{\rm {bd}}/\sigma$ \cite{shiba11,nogu12}.
In this paper, we use $k_{\rm {bend}}=k_{\rm {tilt}}$,
i.e. $C_0  \simeq C_{\rm {bd}}/2\sigma$.

The last term in Eq.~(\ref{eq:U_all}) represents the repulsion between
the different types of molecules
by a monotonic decreasing function:
$U_{\rm {AB}}(r) =  A_1 f_{\rm {cut}}(r)$ with $n=1$, $A_0= 1$, and $r_{\rm {cut}}=2.4\sigma$,
and $A_1=\exp[\sigma/(r_{\rm {cut}}-\sigma)]$ to set $U_{\rm {AB}}(\sigma) = 1$.

\subsection{Simulation Method}

The  $NVT$ ensemble (constant
number of molecules $N$, volume $V$, and temperature $T$) is used
with periodic boundary conditions in a box with side lengths $L_x$, $L_y$, and $L_z$.
A Langevin thermostat is employed to keep the temperature.
The motion of the center of the mass 
${\bf r}^{\rm G}_{i}=({\bf r}^{\rm s}_{i}+{\bf r}^{\rm e}_{i})/2 $ and 
the orientation ${\bf u}_{i}$ are given by underdamped Langevin equations:
\begin{eqnarray} \label{eq:lan1}
  \frac{d {\bf r}^{\rm G}_{i}}{dt} &=& {\bf v}^{\rm G}_{i}, \ \  \frac{d {\bf u}_{i}}{dt} = {\boldsymbol \omega}_{i}, \\
m \frac{d {\bf v}^{\rm G}_{i}}{dt} &=&
 - \zeta_{\rm G} {\bf v}^{\rm G}_{i} + {\bf g}^{\rm G}_{i}(t)
 + {{\bf f}_i}^{\rm G}, \label{eq:lan2}  \\ \label{eq:lan3}
I \frac{d {\boldsymbol \omega}_{i}}{dt} &=&
 - \zeta_{\rm r} {\boldsymbol \omega}_i + ({\bf g}^{\rm r}_{i}(t)
 + {{\bf f}_i}^{\rm r})^{\perp} + \lambda_{\rm L} {\bf u}_{i},
\end{eqnarray}
where $m$ and $I$ denote the mass and the moment of inertia of the molecule, respectively.
The forces are given by ${{\bf f}_i}^{\rm G}= - \partial U/\partial {\bf r}^{\rm G}_{i}$
and ${{\bf f}_i}^{\rm r}= - \partial U/\partial {\bf u}_{i}$ with 
the perpendicular component ${\bf a}^{\perp} ={\bf a}- ({\bf a}\cdot{\bf u}_{i}) {\bf u}_{i}$
and a Lagrange multiplier $\lambda_{\rm L}$ to keep ${\bf u}_{i}^2=1$.
According to  the fluctuation-dissipation theorem,
the friction coefficients $\zeta_{\rm G}$ and $\zeta_{\rm r}$ and 
the Gaussian white noises ${\bf g}^{\rm G}_{i}(t)$ and ${\bf g}^{\rm r}_{i}(t)$
obey the following relations:
the average $\langle g^{\beta_1}_{i,\alpha_1}(t) \rangle  = 0$ and the variance
$\langle g^{\beta_1}_{i,\alpha_1}(t_1) g^{\beta_2}_{j,\alpha_2}(t_2)\rangle  =  
         2 k_{\rm B}T \zeta_{\beta_1} \delta _{ij} \delta _{\alpha_1 \alpha_2} \delta _{\beta_1 \beta_2} \delta(t_1-t_2)$,
where $\alpha_1, \alpha_2 \in \{x,y,z\}$ and  $\beta_1, \beta_2 \in \{{\rm G, r}\}$.

The results are displayed with a length unit of $\sigma$, an energy unit of $k_{\rm B}T$, and
 a time unit of $\tau_0=\zeta_{\rm G}\sigma^2/k_{\rm B}T$.
The Langevin equations are integrated using the leapfrog algorithm \cite{alle87,nogu11}
with $m= \zeta_{\rm G}\tau_0$, $I=\zeta_{\rm r}\tau_0$, $\zeta_{\rm r}=\zeta_{\rm G}\sigma^2$,
and $\Delta t=0.005\tau_0$.

\subsection{Membrane Properties of Single-Component Membranes}
\label{sec:mempro}

The potential-parameter dependence of
the membrane properties of single-component membranes
 (bending rigidity $\kappa$,
area compression modulus, lateral diffusion coefficient, flip-flop rate,
 line tensions $\Gamma_{\rm {edge}}$ of membrane edge, and
 $\Gamma_{\rm {br}}$ of branching junction) 
are investigated in detail in our previous papers~\cite{nogu11,nogu12}.
In this model, these properties can be varied over wide ranges.

The membrane has a wide range of fluid phases, and
the fluid-gel transition point can be controlled by  $\rho^*$.
The area compression modulus $K_{\rm A}$ can be varied by $k_{\rm {tilt}}$.
The flip-flop rate can be varied by $k_{\rm {tilt}}$ and $k_{\rm {bend}}$.
The bending rigidity $\kappa$
is linearly dependent on $k_{\rm {tilt}}$ and $k_{\rm {bend}}$
as  $\kappa/k_{\rm B}T=2.2(k_{\rm {tilt}}+ k_{\rm {bend}})+b_\varepsilon(\varepsilon)$
at $C_{\rm {bd}}=0$.
The line tension $\Gamma_{\rm {edge}}$ of the membrane edge can be controlled by  varying
$\varepsilon$ and $C_{\rm {bd}}$.
As $C_{\rm {bd}}$ increases, 
the line tension $\Gamma_{\rm {edge}}$ decreases.
Consequently, the bilayer membranes become unstable at $\Gamma_{\rm {edge}} \simeq 0$, 
and instead worm-like micelles are formed.

In this paper, 
we use $k_{\rm {bend}}=k_{\rm {tilt}}=4$ or $8$, $\rho^*=14$, and  $C_{\rm bd}^{\rm B}=0$,
so that type B molecules form a fluid membrane with
 $\kappa/k_{\rm B}T=26$ or $44$ for  $\varepsilon_{\rm {att}}^{\rm B}=2$,
respectively.
We use $\varepsilon_{\rm {att}}^{\rm B}=2$  except for Sec. \ref{sec:line_ab}
where  $\varepsilon_{\rm {att}}^{\rm B}$ is varied with $\varepsilon_{\rm {att}}^{\rm A}=\varepsilon_{\rm {att}}^{\rm B}$.
The monolayer (leaflet) of type B membrane has zero spontaneous curvature.
For type A molecules, we use $\varepsilon_{\rm {att}}^{\rm A}=1.5$ or $2$,
and various $C_{\rm bd}^{\rm A}$.
For $\varepsilon_{\rm {att}}^{\rm A}=2$, 
the molecules have a very low critical micelle concentration (CMC);
No or a few isolated molecules are seen in the simulation box.
For  $\varepsilon_{\rm {att}}^{\rm A}=1.5$, 
the molecules have a low but finite value of CMC $\sim 10^{-4}\sigma^{-3}$, 
which corresponds to $0.1$mM for $\sigma \simeq 2$nm.
Vesicles and micelles can change their sizes by absorbing and dissolving type A molecules
 at $\varepsilon_{\rm {att}}^{\rm A}=1.5$.
The simulations with $\varepsilon_{\rm {att}}^{\rm A}=1.5$ and $2$ correspond to
typical lipid-detergent  mixtures and lipid-lipid mixtures, respectively.

The same number of type A and type B molecules are used for simulations of the 
planar membranes in Sec. \ref{sec:plane}. 
In Secs. \ref{sec:line_ab} and  \ref{sec:edge},
$N_{\rm A}=N_{\rm B}=512$ are used.
To investigate membrane pores (Sec. \ref{sec:wet}),
 $N_{\rm A}=N_{\rm B}=1,024$,
$L_x=38\sigma$, $L_y=36\sigma$, $\varepsilon_{\rm {att}}^{\rm A}=\varepsilon_{\rm {att}}^{\rm B}=2$, 
and $k_{\rm {bend}}=8$ are used.
For most simulations of vesicle opening-up,
the  total number of molecule is fixed as $N=4,000$,
and the number ratio is varied. 
In a few cases, $N_{\rm A}=N_{\rm B}=1,000$ are used.
For the opening,
$k_{\rm {bend}}=k_{\rm {tilt}}=8$ and
 $L_x=L_y=L_z=100\sigma$ ($N/V= 0.004/\sigma^3$ for $N=4,000$) 
are used.
The time unit is estimated as $\tau_{\rm 0} \sim 0.1 \mu$s from
the lateral diffusion coefficient $\sim 10^{-8}$cm$^2$/s for phospholipids \cite{wu77}.
We consider that micelles 
 maintaining their shapes for an  interval of $100,000\tau_0$ ($\sim 10$ms) 
are in steady (meta)stable states.
The error bars of data are estimated 
 from three simulation runs.

\begin{figure}
\includegraphics{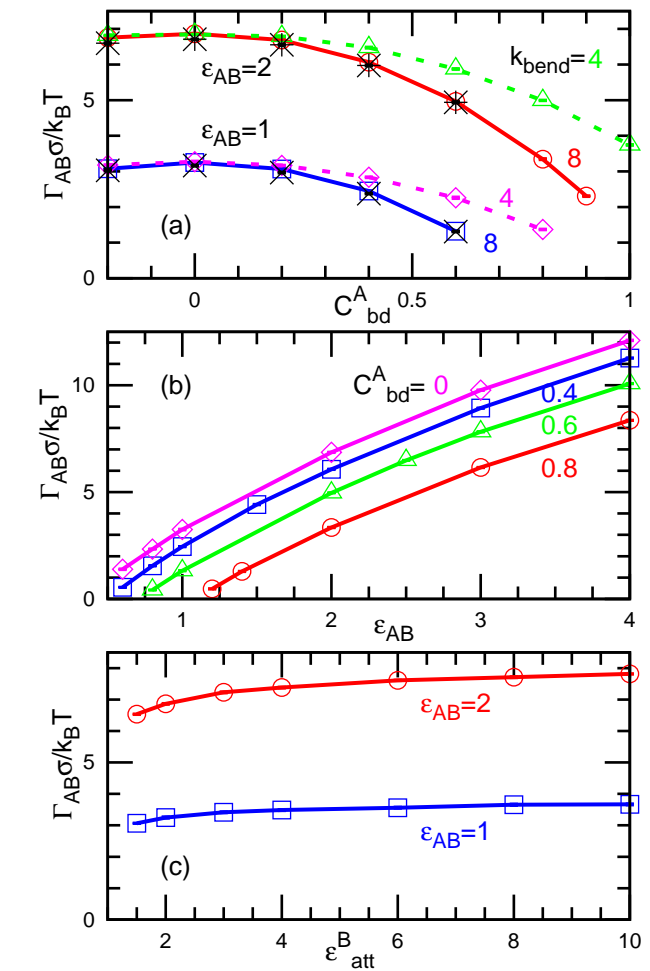}
\caption{\label{fig:g_ab}
Line tension $\Gamma_{\rm {AB}}$ of the phase boundary of membrane domains
in tensionless membranes.
(a) Dependence on $C_{\rm {bd}}^{\rm A}$.
The dashed and solid lines represent data for $k_{\rm {bend}}=4$ and $8$, 
respectively.
The symbols ($\Box, \diamond$)  and ($\circ,\triangle$) represent data 
for $\varepsilon_{\rm {AB}}=1$ and $2$, respectively,
at $\varepsilon_{\rm {att}}^{\rm A}=\varepsilon_{\rm {att}}^{\rm B}=2$.
The symbols ($\times$) and ($\ast$) represent data 
for $\varepsilon_{\rm {AB}}=1$ and $2$, respectively, 
at  $\varepsilon_{\rm {att}}^{\rm A}=1.5$, $\varepsilon_{\rm {att}}^{\rm B}=2$, 
and $k_{\rm {bend}}=8$.
(b) Dependence on $\varepsilon_{\rm {AB}}$
at  $k_{\rm {bend}}=8$,  $\varepsilon_{\rm {att}}^{\rm A}=\varepsilon_{\rm {att}}^{\rm B}=2$,
 and  $C_{\rm {bd}}^{\rm A}=0$ ($\diamond$), $0.4$ ($\Box$), $0.6$ ($\triangle$), or $0.8$ ($\circ$).
(c) $\Gamma_{\rm {AB}}$ for various $\varepsilon_{\rm {att}}^{\rm B}$
at  $\varepsilon_{\rm {att}}^{\rm A}=\varepsilon_{\rm {att}}^{\rm B}$,
$k_{\rm {bend}}=8$,  $C_{\rm {bd}}^{\rm A}=0$, and $\varepsilon_{\rm {AB}}=1$ ($\Box$) or $2$ ($\circ$).
 The error bars are smaller than the size of the symbols.
}
\end{figure}

\begin{figure}
\includegraphics{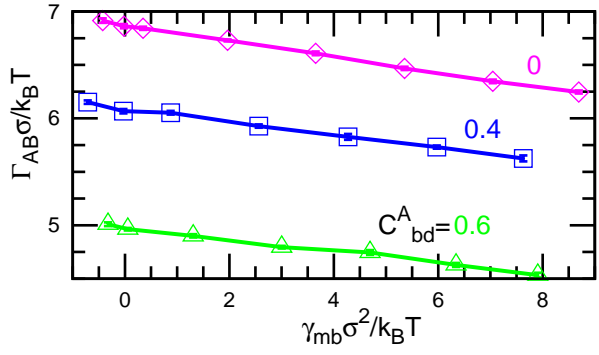}
\caption{\label{fig:g_ten}
Surface tension $\gamma$  dependence of
the line tension  $\Gamma_{\rm {AB}}$ of the phase boundary
at  $\varepsilon_{\rm {AB}}=2$,  $\varepsilon_{\rm {att}}^{\rm A}=\varepsilon_{\rm {att}}^{\rm B}=2$, 
$k_{\rm {bend}}=8$, and  $C_{\rm {bd}}^{\rm A}=0$ ($\diamond$), $0.4$ ($\Box$), or $0.6$ ($\triangle$).
 The error bars are smaller than the size of the symbols.
}
\end{figure}

\section{Planar Membranes}
\label{sec:plane}

\subsection{Line Tension of Phase Boundary}
\label{sec:line_ab}

First, we estimate the line tension $\Gamma_{\rm {AB}}$ of the boundary between
the type A and B membrane domains.
Two domains are prepared
on the planar membrane along the $xy$ plane, and 
two lines of the phase  boundary are set parallel to the $x$ axis.
The  line tension $\Gamma_{\rm {AB}}$  is given by \cite{tolp04,lara05,reyn08}
\begin{equation}
\Gamma_{\rm {AB}} = \langle P_{yy} - P_{xx} \rangle V/2L_x,
\label{eq:stpt}
\end{equation} 
where $V=L_xL_yL_z$ and $2L_x$ is the total boundary length.
The diagonal components of the pressure tensor are given by
\begin{equation}
P_{\alpha\alpha} = (Nk_{\rm B}T - 
     \sum_{i} \alpha_{i}\frac{\partial U}{\partial {\alpha}_{i}} )/V,
\end{equation} 
where  $\alpha \in \{x,y,z\}$.
We  calculated $\Gamma_{\rm {AB}}$ at
three length ratios $L_y/L_x=1.8$, $2$, and $2.2$;
and we checked that their differences were less than the statistical error bars.
The membrane surface tension $\gamma_{\rm {mb}}$ is given by
$\gamma_{\rm {mb}}= -\langle P_{yy} \rangle L_z$,
since  $P_{zz} \simeq 0$ for
solvent-free simulations with low CMC.

The line tension $\Gamma_{\rm {AB}}$ can be controlled by $\varepsilon_{\rm {AB}}$ (see Fig. \ref{fig:g_ab}).
It  increases with increasing $\varepsilon_{\rm {AB}}$ and
decreases with increasing $C_{\rm {bd}}^{\rm A}$.
The $C_{\rm {bd}}^{\rm A}$ dependence is qualitatively similar to that of the line tension  $\Gamma_{\rm {edge}}$
of the membrane edges.
On the other hand, the $\varepsilon_{\rm {att}}$ dependence is different;
$\Gamma_{\rm {AB}}$ is almost independent of  $\varepsilon_{\rm {att}}$ [see Fig. \ref{fig:g_ab}(c)], 
while  $\Gamma_{\rm {edge}}$ increases with increasing  $\varepsilon^{\rm {B}}_{\rm {att}}$.\cite{nogu11}

The tension  $\Gamma_{\rm {AB}}$ slightly decreases with increasing surface tension $\gamma_{\rm {mb}}$
as shown in Fig. \ref{fig:g_ten}.
The surface tension $\gamma_{\rm {mb}}$ is varied by changing the projected membrane area in the $xy$ plane.
The repulsive interaction between different types of molecules
is likely  to be reduced by the slightly larger distances of neighboring molecules at larger $\gamma_{\rm {mb}}$.

\begin{figure}
\includegraphics{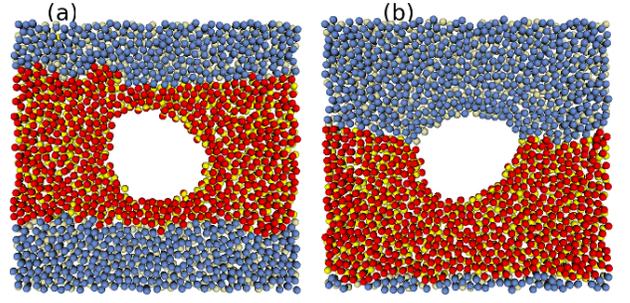}
\caption{\label{fig:snap_pore}
Snapshots of a bilayer membrane with a pore
at $\varepsilon_{\rm {att}}^{\rm A}=\varepsilon_{\rm {att}}^{\rm B}=2$, 
$k_{\rm {bend}}=8$, and  $C_{\rm {bd}}^{\rm A}=0.4$.
(a) $\varepsilon_{\rm {AB}}=1$. (b)  $\varepsilon_{\rm {AB}}=2$.
The red (yellow) and light blue (light yellow) hemispheres represent the hydrophilic 
(hydrophobic) parts of type A  and type B molecules,
respectively.  
}
\end{figure}

\begin{figure}
\includegraphics{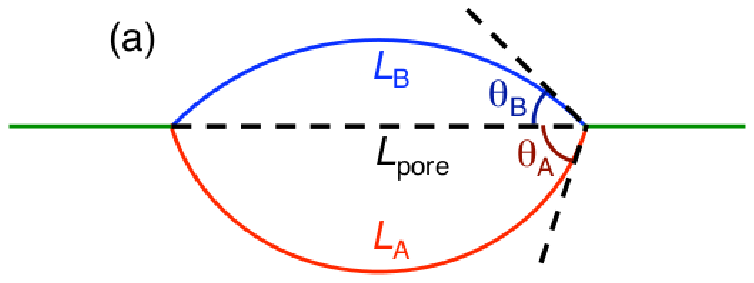}
\includegraphics{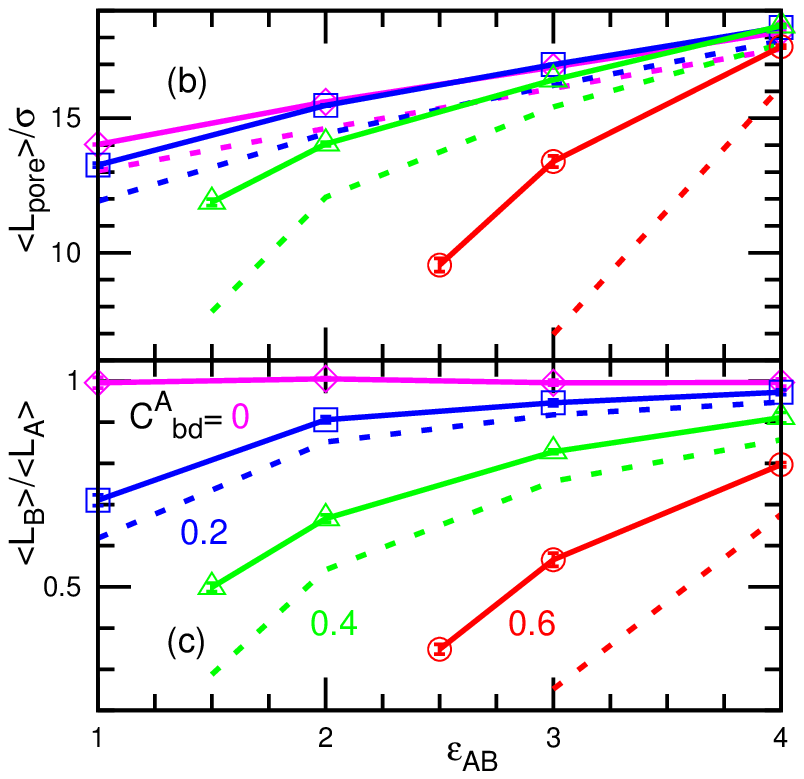}
\caption{\label{fig:wet}
(a) Schematic representation of a pore on the phase boundary.
(b) Pore transverse diameter $L_{\rm {pore}}$ and (c) the ratio $L_{\rm {B}}/L_{\rm {A}}$ of  edge  lengths 
at  $\varepsilon_{\rm {att}}^{\rm A}=\varepsilon_{\rm {att}}^{\rm B}=2$, 
$k_{\rm {bend}}=8$, and  $C_{\rm {bd}}^{\rm A}=0$ ($\diamond$), $0.2$ ($\Box$), $0.4$ ($\triangle$), or $0.6$ ($\circ$).
The solid lines with the symbols represent simulation data.
The dashed lines are obtained from Eq. (\ref{eq:lpore}).
 The error bars are smaller than the size of the symbols.
}
\end{figure}

\subsection{Pore on Phase Boundary}
\label{sec:wet}

As the membrane area increases, increasing surface tension induces a pore opening in the membrane.
In the phase-separated membrane, the pore stays on the phase boundary 
or in the domain with lower $\Gamma_{\rm {edge}}$ (see Fig. \ref{fig:snap_pore}). 
Here, we investigate the pore position and shape in the phase-separated membranes
for various  $C_{\rm {bd}}^{\rm A}$ and $\varepsilon_{\rm {AB}}$.

The pore transverse diameter $L_{\rm {pore}}$ and 
 edge length ratio $L_{\rm {B}}/L_{\rm {A}}$ of  type A and type B membranes
 are estimated as follows.
First, the pore area is calculated using a particle-insertion method \cite{tolp04,nogu06}.
In the $xy$ plane, 
$10,000$ ghost particles with a hard-sphere diameter $\sigma$ are distributed 
randomly, and then the particle placement is accepted if
no overlap with the projections of the hard cores of the membrane 
particles occurs. 
Next, the molecules closer than  $1.5\sigma$ to any  ghost pore particle
are considered to belong to the membrane edge.
The neighbor distance between molecules are assumed to be 
same for the both domains, so
the length ratio is given by the number ratio of molecules on the edge: 
$L_{\rm {B}}/L_{\rm {A}}= N_{\rm {pore}}^{\rm {B}}/N_{\rm {pore}}^{\rm {A}}$.
Two mutual contact points of the domain boundary and membrane edges
are defined as the middle points between the nearest neighbor edge molecules of 
types A and B [see  Fig. \ref{fig:wet}(a)].
The  transverse diameter $L_{\rm {pore}}$  is given by the length between the two contact points along the $x$ axis.
Isolated ghost particles and molecules are not taken into account in the above calculations.

\begin{figure}
\includegraphics{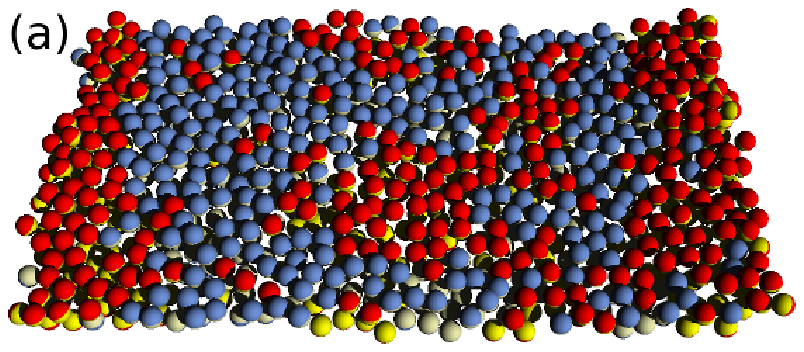}
\includegraphics{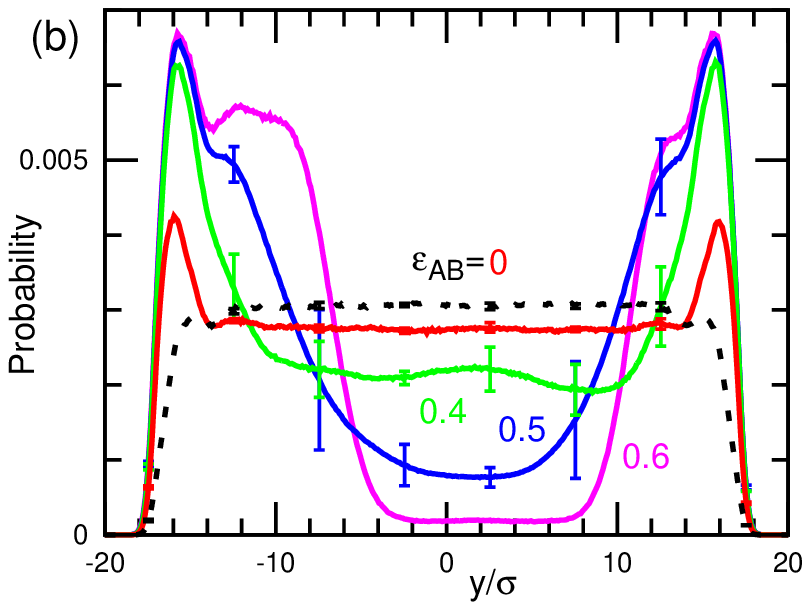}
\caption{\label{fig:hisy}
Bilayer membrane strip at  $\varepsilon_{\rm {att}}^{\rm A}=2$, 
$k_{\rm {bend}}=8$, and  $C_{\rm {bd}}^{\rm A}=0.4$.
(a) Snapshot of partially phase-separated membranes
at $\varepsilon_{\rm {AB}}=0.4$.
(b) Probability distribution in the $y$ direction, which is taken along the eigenvector of 
the maximum eigenvalue of the $2$D gyration tensor.
The solid lines represent data for the type A at $\varepsilon_{\rm {AB}}=0$, $0.4$, $0.5$, and $0.6$.
The dashed line represents data for the type B at  $\varepsilon_{\rm {AB}}=0$.
The error bars are shown at several data points for $\varepsilon_{\rm {AB}}=0$, $0.4$, and $0.5$.
For $\varepsilon_{\rm {AB}}=0.6$, the distribution calculated from a single run
is shown.
}
\end{figure}

At $C_{\rm {bd}}^{\rm A}=0$, the two domains are identical.
Thus, the average pore shape is symmetric, and
the edge lengths $\langle L_{\rm A} \rangle$ and $\langle L_{\rm B} \rangle$ of type A and type B membranes
are equal (see Fig. \ref{fig:wet}). 
As $C_{\rm {bd}}^{\rm A}$ increases,
the center of the pore moves to the type A domain.
As $\varepsilon_{\rm {AB}}$ increases,
the pore elongates along the phase boundary,
and  $\langle L_{\rm A} \rangle/\langle L_{\rm B} \rangle$ approaches unity.

\begin{figure}
\includegraphics{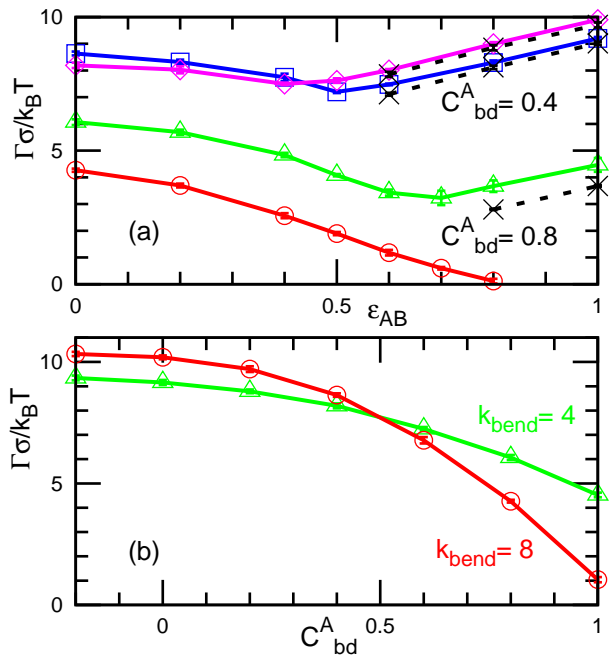}
\caption{\label{fig:g_edge}
Effective line tension $\Gamma$ of the membrane edge
at  $\varepsilon_{\rm {att}}^{\rm A}=2$.
(a) The  squares ($\Box$) and circles ($\circ$) represent data for 
 $C_{\rm {bd}}^{\rm A}=0.4$ and $0.8$, respectively, at
$k_{\rm {bend}}=8$.
The diamonds  ($\diamond$) and triangles ($\triangle$) represent data for 
 $C_{\rm {bd}}^{\rm A}=0.4$ and $0.8$, respectively, at
$k_{\rm {bend}}=4$.
The dashed lines with ($\times$) represent 
 $\Gamma=\Gamma_{\rm {A}}+\Gamma_{\rm {AB}}$.
(b)  The triangles ($\triangle$) and  circles ($\circ$) represent data for 
$k_{\rm {bend}}=4$ and $8$, respectively, at $C_{\rm {bd}}^{\rm A}=0$.
 The error bars are smaller than the size of the symbols.
}
\end{figure}

These pore shapes are understood by the Young equation and Laplace pressure.
Three line tensions are balanced at the mutual contact points. \cite{safr94,dege03}
\begin{eqnarray}\label{eq:young}
\Gamma_{\rm {AB}} &=& \Gamma_{\rm {A}}\cos(\theta_{\rm A}) + \Gamma_{\rm {B}}\cos(\theta_{\rm B}), \nonumber \\
\Gamma_{\rm {A}}\sin(\theta_{\rm A}) &=& \Gamma_{\rm {B}}\sin(\theta_{\rm B}),
\end{eqnarray}
where $\Gamma_{\rm {A}}$ and $\Gamma_{\rm {B}}$ denote the line tensions of the membrane edges of 
the type A and type B domains,
respectively [see  Fig. \ref{fig:wet}(a)].
Therefore, the contact angle $\theta_{\rm A}$ is written as
\begin{equation} 
\cos(\theta_{\rm A}) = \frac{\Gamma_{\rm {AB}}^2+\Gamma_{\rm {A}}^2-\Gamma_{\rm {B}}^2}{2\Gamma_{\rm {A}}\Gamma_{\rm {AB}}}.
\end{equation}\label{eq:angle}
The membrane edges of the domains have constant curvature radii $r_{\rm A}$ and $r_{\rm B}$.
The edge line tension is balanced with the two-dimensional Laplace pressure  $\Delta\gamma$: 
$\Gamma_{\rm {A}}= r_{\rm A}\Delta\gamma$ and  $\Gamma_{\rm {B}}= r_{\rm B}\Delta\gamma$.
The transverse diameter $L_{\rm {pore}}$ and 
 edge lengths of each domain are given by 
\begin{eqnarray}
L_{\rm {pore}} &=& 2r_{\rm A}\sin(\theta_{\rm A})=2r_{\rm B}\sin(\theta_{\rm B}), \nonumber \\
L_{\rm {A}} &=& 2r_{\rm A}\theta_{\rm A}, \ \ L_{\rm {B}}=2r_{\rm B}\theta_{\rm B}.
\label{eq:lpore}
\end{eqnarray}
Thus, the pore lengths are obtained from the line tensions and the surface tension difference $\Delta\gamma$.
Since the pore has zero surface tension,  $\Delta\gamma$ is calculated as
$\Delta\gamma=\gamma_{\rm {mb}}= -\langle P_{yy} \rangle L_z$.
These analytical results qualitatively explain the simulation results (see Fig. \ref{fig:wet}). 
However, $L_{\rm {pore}}$ and $L_{\rm {B}}/L_{\rm {A}}$ are underestimated in particular 
for the parameter sets close to the wetting transition points.
These deviations are likely caused by the negligence of the thermal fluctuations in the theory 
or the finite size effects in the simulations.

In order to form bicelles with membrane edges covered by type A molecules,
the complete wetting condition $\Gamma_{\rm {B}} > \Gamma_{\rm {AB}}+ \Gamma_{\rm {A}}$
is necessary: $\varepsilon_{\rm {AB}} \leq 1$ at $C_{\rm {bd}}^{\rm A}=0.4$ and  
$\varepsilon_{\rm {AB}} \leq 2$ at $C_{\rm {bd}}^{\rm A}=0.6$.
At $C_{\rm {bd}}^{\rm A}=0.8$, a pore is unstable and worm-like micelles grow from the edge of the type A domain.

\subsection{Edge Effects on Phase Separation}
\label{sec:edge}

Next, we consider a membrane strip, which has two open edges parallel to the $x$ axis [see Fig. \ref{fig:hisy}(a)].
Since the type A domain has low line tension $\Gamma_{\rm {A}}<\Gamma_{\rm {B}}$ at $C_{\rm {bd}}^{\rm A}>0$,
the type A molecules prefer to remain at the membrane edges.
The probability of type A molecules is higher than that of the type B at the membrane edges 
even for $\varepsilon_{\rm {AB}}=0$ 
[compare solid and dashed lines for $\varepsilon_{\rm {AB}}=0$ in Fig. \ref{fig:hisy}(b)].
The free-energy reduction of the edges is balanced with the mixing entropy of the two species.
Similar concentrating effects were reported in the cases of the two-component vesicles \cite{gozd06} 
and a vesicle with a cylindrical micelle  \cite{gree11}, where
 components with higher spontaneous curvature are concentrated at the high curvature regions.

The effective line tension $\Gamma$ of the membrane edge
is estimated as 
\begin{equation}
\Gamma = -\langle P_{xx} \rangle V/2L_x.
\label{eq:g_eff}
\end{equation} 
For single-component membranes, $\Gamma$ gives the line tension of the  membrane edges.
For the phase-separated membranes, $\Gamma$ gives the sum of the two line tensions: 
$\Gamma=\Gamma_{\rm {A}}+\Gamma_{\rm {AB}}$ [compare dashed and solid lines in Fig. \ref{fig:g_edge}(a)].
Surprisingly, $\Gamma$ decreases with increasing  $\varepsilon_{\rm {AB}}$
at low $\varepsilon_{\rm {AB}}$,
where the two surfactants are in the mixing condition (see Fig. \ref{fig:hisy}).
This decrease is caused by the coverage of the membrane edges by type A molecules,
which reduces the free energy of the edges.
At the phase separation threshold
 $\Gamma$ is minimum, and it increases with increasing $\varepsilon_{\rm {AB}}$
for completely phase-separated membranes.
At $\varepsilon_{\rm {AB}}=0$, $\Gamma$ is slightly lower than $\Gamma_{\rm {edge}}$ for the mean spontaneous curvature
$C_{\rm {bd}}=(C_{\rm {bd}}^{\rm A}+C_{\rm {bd}}^{\rm B})/2$,
because of the high concentration of type A molecules at the edges [see Fig. \ref{fig:g_edge}(b)].

\begin{figure}
\includegraphics{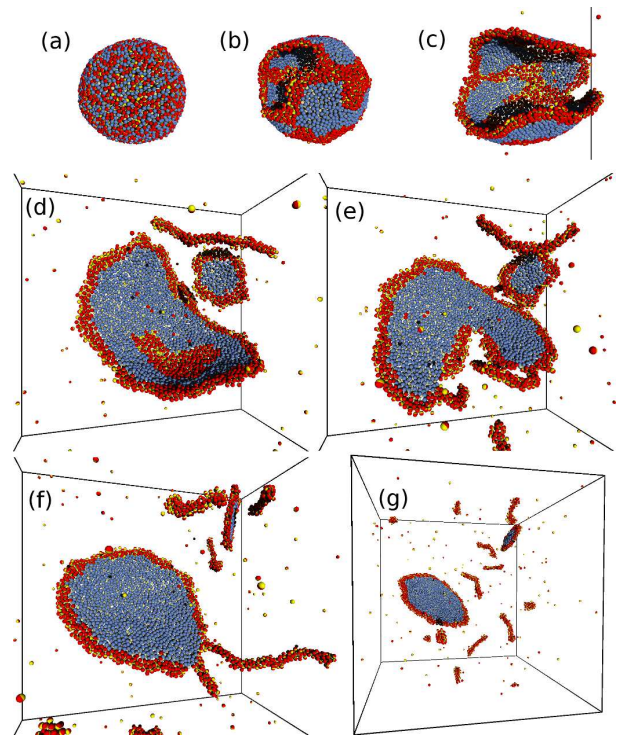}
\caption{\label{fig:snaprup15}
Sequential snapshots of bicelle formation from a vesicle
at $N_{\rm A}=N_{\rm B}=2,000$, $\varepsilon_{\rm {att}}^{\rm A}=1.5$,
 $\varepsilon_{\rm {AB}}=1$, and $C_{\rm {bd}}^{\rm A}=0.8$.
(a) $t=0$. (b) $t=1,000\tau_0$. (c) $t=1,200\tau_0$. (d) $t=4,000\tau_0$.
(e) $t=5,000\tau_0$. (f) $t=12,000\tau_0$. (g) $t=28,000\tau_0$.
The snapshots in (a-f) are seen from the same viewpoint.
}
\end{figure}

\begin{figure}
\includegraphics{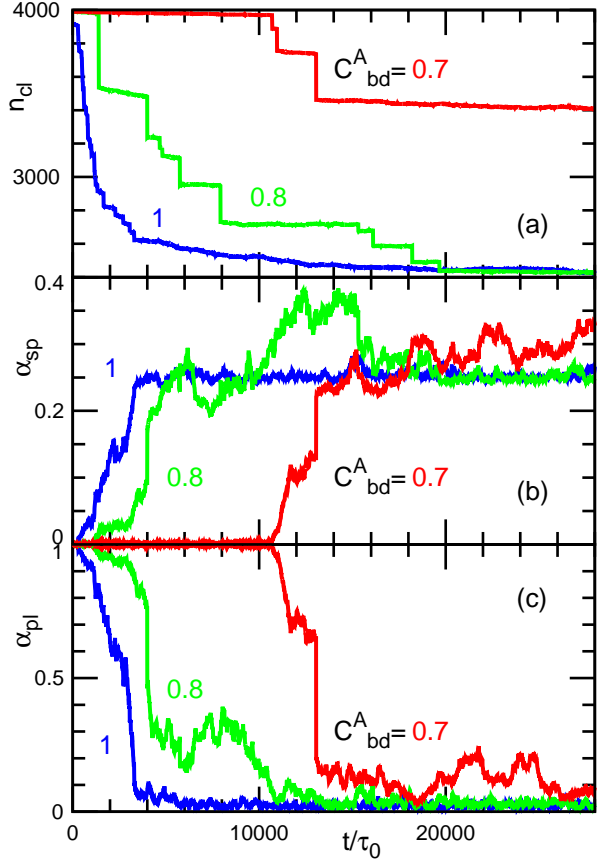}
\caption{\label{fig:rupte15}
Time development of the bicelle formation
at $N_{\rm A}=N_{\rm B}=2,000$, $\varepsilon_{\rm {att}}^{\rm A}=1.5$,
 $\varepsilon_{\rm {AB}}=1$, and $C_{\rm {bd}}^{\rm A}=0.7$, $0.8$, or $1$.
(a) The number of molecules $n_{\rm {cl}}$, (b) asphericity $\alpha_{\rm {sp}}$,
and (c) aplanality $\alpha_{\rm {pl}}$ of the largest cluster.
The same data as in Fig. \ref{fig:snaprup15} are shown for $C_{\rm {bd}}^{\rm A}=0.8$.
}
\end{figure}

\section{Vesicle Opening-Up}
\label{sec:openup}

As the spontaneous curvature of the monolayer is increased 
by changing $C_{\rm {bd}}$ ($C_0  \simeq C_{\rm {bd}}/2\sigma$),
 single-component vesicles are ruptured, and these vesicles subsequently
transform into branched worm-like micelles \cite{nogu11}.
For two-component vesicles, the number ratio of the components
and the line tension $\Gamma_{\rm {AB}}$ of the phase boundary 
are also key parameters.
First, we describe the vesicle opening-up for $\varepsilon_{\rm {att}}^{\rm A}=1.5$,
where type A molecules have finite CMC.
The vesicle rupture occurs  by changing $C^{\rm A}_{\rm {bd}}$ or $\Gamma_{\rm {AB}}$.
Next, we describe the case for $\varepsilon_{\rm {att}}^{\rm A}=2$,
where both types of molecules have negligibly small CMC.
We call the former and latter mixtures 
lipid-detergent mixtures and lipid-lipid mixtures, respectively.

\begin{figure}
\includegraphics{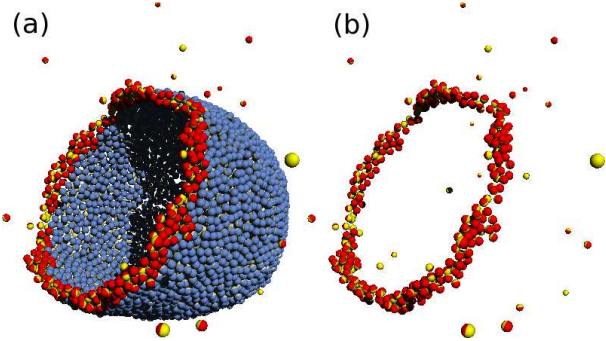}
\caption{\label{fig:snapcup}
Snapshots of a cup-shaped vesicle
at $N_{\rm A}=400$, $N_{\rm B}=3,600$, $\varepsilon_{\rm {att}}^{\rm A}=1.5$,
 $\varepsilon_{\rm {AB}}=1$, and $C_{\rm {bd}}^{\rm A}=0.8$.
All molecules and type A molecules are shown in (a) and (b), respectively.
}
\end{figure}

\begin{figure}
\includegraphics{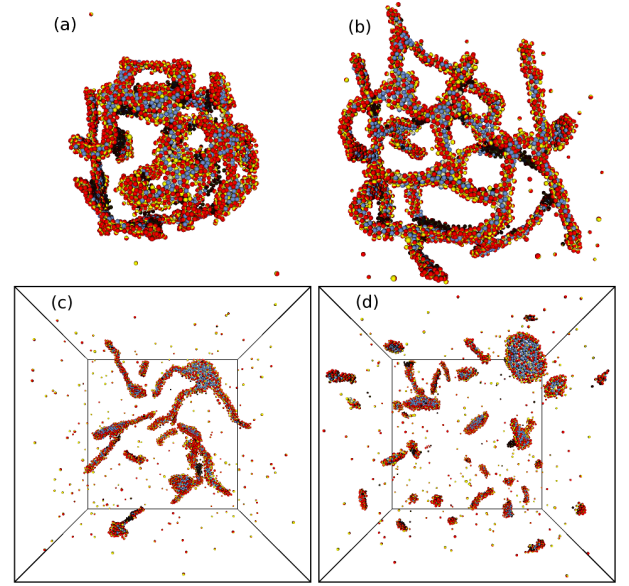}
\caption{\label{fig:snape15d3}
Sequential snapshots of bicelle formation from a vesicle
at $N_{\rm A}=3,000$, $N_{\rm B}=1,000$, $\varepsilon_{\rm {att}}^{\rm A}=1.5$,
 $\varepsilon_{\rm {AB}}=0$, and $C_{\rm {bd}}^{\rm A}=1$.
(a) $t=100\tau_0$. (b) $t=1,000\tau_0$. (c) $t=20,000\tau_0$. (d) $t=100,000\tau_0$.
All snapshots are seen from the same viewpoint.
}
\end{figure}

\subsection{Lipid-Detergent Mixtures}
\label{sec:lipdet}

Typical examples of bicelle formation in lipid-detergent mixtures
at $N_{\rm A}/N_{\rm B}=1$
are shown in Figs. \ref{fig:snaprup15} and \ref{fig:rupte15}.
The vesicle is equalibrated at  $\varepsilon_{\rm {AB}}=0$
and subsequently $\varepsilon_{\rm {AB}}$ is suddenly changed to $\varepsilon_{\rm {AB}}=1$ at $t=0$.
After the change, the molecules form domains, and then a pore opens in a type A domain 
[see Figs. \ref{fig:snaprup15}(a), (b)].
The pore grows along the type A domain and the vesicle opens up [see Figs. \ref{fig:snaprup15}(c), (d)].
In Fig. \ref{fig:snaprup15}(d),
a bicelle and a worm-like micelle are detached from the largest cluster.
The type A domain within the large bicelle diffuses to the membrane edge
and fuses the other type A domain surrounding the type B bilayer disk [see Fig. \ref{fig:snaprup15}(e)].
As consequence of the domain fusion, two worm-like micelles connected with the bicelle are formed
[see Fig. \ref{fig:snaprup15}(f)].
After a while, the worm-like micelles are detached, and a large discoidal bicelle is formed [see Fig. \ref{fig:snaprup15}(g)].
A movie for another example of the opening dynamics at  $C_{\rm {bd}}=0.8$ is provided in ESI (Movie 1).
In the movie, closure of a small pore occurs inside a bilayer disk during the bicelle formation.

At larger $C_{\rm {bd}}$, the vesicle opening begins earlier in average,
since a pore can open in a smaller domain.
The connections between bicelles and worm-like micelles have longer lifetimes at smaller  $C_{\rm {bd}}$.
The time development of the largest clusters is shown in Fig. \ref{fig:rupte15}.
The $i$-th and $j$-th molecules are considered to belong to the same cluster at $r_{ij}^{\rm s}<r_{\rm {att}}$.
The number of molecules $n_{\rm {cl}}$ in the largest cluster decreases in a stepwise manner 
as the micelles undergo detachment.
The shape of the clusters can be characterized by the ratios of the three eigenvalues
${\lambda_1}, {\lambda_2}, {\lambda_3}$
 of the gyration tensors
$a_{\alpha\beta}= (1/n_{\rm {cl}})\sum_i (\alpha^{\rm s}_{i}-\alpha_{\rm G})
(\beta^{\rm s}_{i}-\beta_{\rm G})$,
where the center of the mass $\alpha_{\rm G} = (1/n_{\rm {cl}})\sum_i \alpha^{\rm s}_{i}$ and
$\alpha, \beta \in x,y,z$.
The asphericity $\alpha_{\rm {sp}}$ \cite{rudn86} and aplanarity $\alpha_{\rm {pl}}$ \cite{nogu06}
are defined as
\begin{eqnarray}
\alpha_{\rm {sp}} &=&  \frac{({\lambda_1}-{\lambda_2})^2 + 
  ({\lambda_2}-{\lambda_3})^2+({\lambda_3}-{\lambda_1})^2}{2 (\lambda_1 + \lambda_2 + \lambda_3)^2}, \\
\alpha_{\rm {pl}} &=& \frac{9\lambda_1\lambda_2\lambda_3} {(\lambda_1+\lambda_2+\lambda_3)
    (\lambda_1\lambda_2+\lambda_2\lambda_3+\lambda_3\lambda_1)},
\end{eqnarray}
to quantify
 the degrees of deviation from a spherical 
shape and planar shape, respectively.
These two quantities are  convenient measures to distinguish spherical, discoidal, and cylindrical shapes: 
 $(\alpha_{\rm {sp}}, \alpha_{\rm {pl}})=(0,1)$ for spheres, 
$(\alpha_{\rm {sp}}, \alpha_{\rm {pl}})=(1,0)$ for thin rods, and
 $(\alpha_{\rm {sp}}, \alpha_{\rm {pl}})=(0.25,0)$ for thin disks.
As a vesicle opens up, $\alpha_{\rm {sp}}$ and $\alpha_{\rm {pl}}$
show a rapid increase and decrease, respectively.
After the worm-like micelles are detached, 
the bicelle has a disk shape, $(\alpha_{\rm {sp}}, \alpha_{\rm {pl}}) \simeq (0.25,0)$.

\begin{figure}
\includegraphics{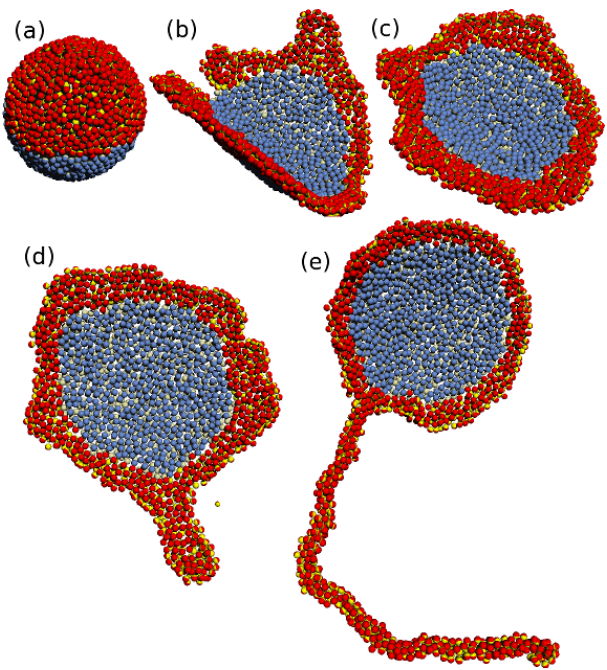}
\caption{\label{fig:snapen2ke2d1}
Sequential snapshots of a racket-like bicelle formation 
at $N_{\rm A}=N_{\rm B}=1,000$, $\varepsilon_{\rm {att}}^{\rm A}=2$,
 $\varepsilon_{\rm {AB}}=2$, and $C_{\rm {bd}}^{\rm A}=0.8$.
(a) $t=8,500\tau_0$. (b) $t=9,000\tau_0$. (c) $t=10,000\tau_0$. (d) $t=48,000\tau_0$.
(e) $t=60,000\tau_0$.
}
\end{figure}

\begin{figure}
\includegraphics{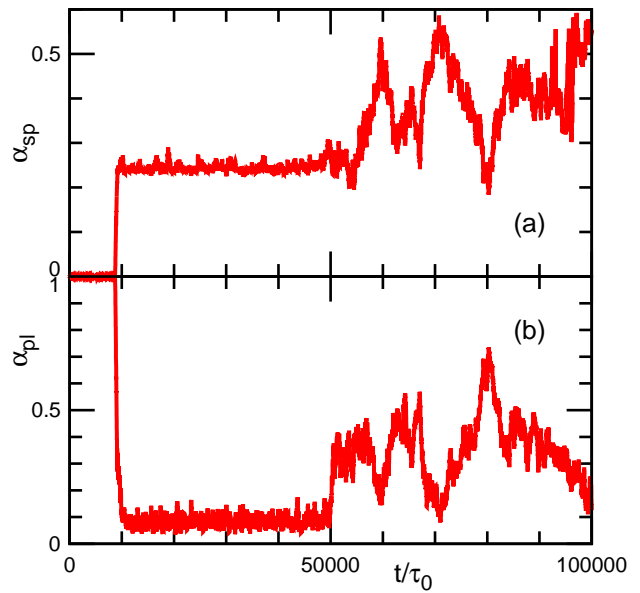}
\caption{\label{fig:rupte2}
Time development of (a) asphericity $\alpha_{\rm {sp}}$
and (b) aplanality $\alpha_{\rm {pl}}$ 
at $N_{\rm A}=N_{\rm B}=1,000$, $\varepsilon_{\rm {att}}^{\rm A}=2$,
 $\varepsilon_{\rm {AB}}=2$, and $C_{\rm {bd}}^{\rm A}=0.8$.
The same data as in Fig. \ref{fig:snapen2ke2d1} are shown.
}
\end{figure}

When type A molecules are too few to cover the circumference of the bilayer disk,
a cup-like vesicle can be formed.
Figure \ref{fig:snapcup} shows a stable cup-like vesicle.
The chemical potential of type A molecules surrounding the rim
is balanced with the bulk chemical potential.
Similar stable cup-like vesicles are experimentally observed 
for liposomes with talins. \cite{sait98}

When vesicles contain many type A molecules,
the vesicles are ruptured even at $\varepsilon_{\rm {AB}}=0$.
At  $N_{\rm A}/N_{\rm B}=3$,
the rupture begins 
as soon as a sufficiently high spontaneous curvature is added by changing $C_{\rm {bd}}^{\rm A}=0$ to $1$
(see Fig. \ref{fig:snape15d3}).
First, a branched worm-like micelle is formed like in the single-component case.
Then, however, type B molecules assemble into the branching junctions
and form bilayer membranes unlike in the single-component case
[see a bicelle with three arms in the top right of Fig. \ref{fig:snape15d3}(c)].
Finally,  bicelles and worm-like micelles without branches are formed.
As discussed in Sec. \ref{sec:edge}, 
inhomogeneous curvature inside of micelles induces phase segregation.
Furthermore, the segregation induces morphological changes of micelles.
Thus, the phase segregation and morphological changes can be coupled
in micelle formation.

\subsection{Lipid-Lipid Mixtures}
\label{sec:liplip}

In lipid-lipid mixtures,
the number of molecules in surfactant aggregates are changed 
principally only via fission or fusion of aggregates,
since the both surfactants have negligibly small CMC.
Figures \ref{fig:snapen2ke2d1} and \ref{fig:rupte2} show the opening dynamics from a phase-separated vesicle
at $N=2,000$ and $C_{\rm {bd}}^{\rm A}=0.8$. 
First, the opened membrane forms a bicelle.
This bicelle maintains its shape for $38,000\tau_0$.
Then, 
a worm-like micelle grows from a bicelle edge,
and a racket-like micelle is formed [see Figs. \ref{fig:snapen2ke2d1}(d), (e)].
The excess type A molecules in the bicelle rim move to the worm-like micelle.
The bicelle and racket-like micelle can be distinguished by  $\alpha_{\rm {sp}}$ and $\alpha_{\rm {pl}}$
(see Fig. \ref{fig:rupte2}).
The worm-like micelle is not detached unlike in the finite CMC simulation at $\varepsilon_{\rm {att}}^{\rm A}=1.5$.
We checked the racket shape is maintained for $100,000\tau_0$.  
At $N_{\rm A}=N_{\rm B}=2,000$ and $C_{\rm {bd}}^{\rm A}=0.8$, the vesicles are not ruptured.
Since higher curvature of smaller vesicles slightly reduces a free-energy barrier of the pore opening,
the vesicle rupture  occurs more frequently in smaller vesicles.

\begin{figure}
\includegraphics{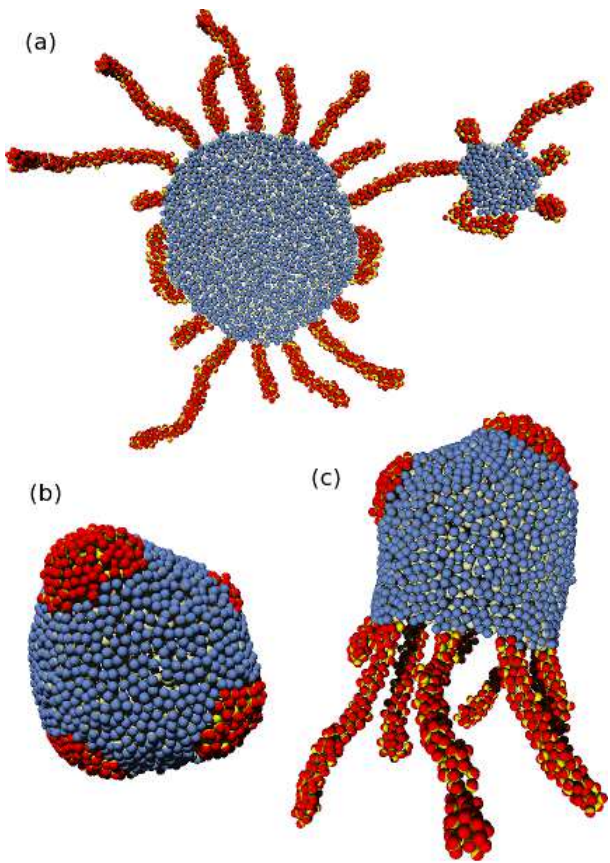}
\caption{\label{fig:snape2}
Snapshots of various obtained morphologies of surfactant aggregates
at $N_{\rm A}=N_{\rm B}=2,000$, $\varepsilon_{\rm {att}}^{\rm A}=2$,
 $\varepsilon_{\rm {AB}}=2$, and $C_{\rm {bd}}^{\rm A}=1$.
}
\end{figure}

For vesicles with a high spontaneous curvature at $C_{\rm {bd}}^{\rm A}=1$,
ruptured membranes transform into bicelles connected with  several worm-like micelles (see Fig. \ref{fig:snape2}).
Movies 2 and 3 in ESI show the formation dynamics
of the octopus-like micelles in Figs. \ref{fig:snape2}(a) and (c), respectively. 
After the change from $\varepsilon_{\rm {AB}}=0$ to $2$,
type A molecules assemble to form domains.
As the domains become sufficiently large,
the membranes in the type A domains are ruptured  and
the excess molecules form connected worm-like micelles.
These processes are similar to those in lipid-detergent mixtures shown in Fig. \ref{fig:snaprup15}.
However, the later processes are different.
Type A molecules do not form a circular rim surrounding type B domains.
Instead, worm-like micelles are directly connected to the bilayer disk of the type B domains
as shown in Fig. \ref{fig:snape2}(a).
These  worm-like micelles are stable and their lengths are fixed in long simulation runs.
At the side of the connection junctions, the membrane edges of the type B domain
are locally exposed in vapor (solvent) space. 

When multiple pores are opened at the same time,
small pores are closed and the worm-like micelles on the pores are detached.
If all the pores are closed, polyhedral vesicles are formed [see Fig. \ref{fig:snape2}(b)].
The asymmetric molecular distribution of the inner and outer leaflets can generate 
a spontaneous curvature of the bilayer in the type A domains.
Hence, the type A domains remain at the vertices with higher curvatures.
The formation of similar polyhedral vesicles has been reported in experiments \cite{veat03,gudh07} and other simulations \cite{huan11,hu11}.
A large pore often remains and
 a cup-like vesicle connected with several worm-like micelles is formed 
as a steady state [see Fig. \ref{fig:snape2}(c) and Movie 3].

When the spontaneous curvature of the octopus-like micelles is changed from $C_{\rm {bd}}^{\rm A}=1$ to $0.8$,
type A molecules form the rim of the bilayer disk
connected with worm-like micelles  like the micelle shown in Fig. \ref{fig:snapen2ke2d1}.
The number of connected worm-like micelles decreases gradually because of the length fluctuations due to thermal diffusion.
Type A molecules in removed short micelles move to remaining long micelles.
a few long worm-like micelles still exist after long simulation runs.
The cup-like vesicles with worm-like micelles  are opened up into flat bicelles.
When the value of $C_{\rm {bd}}^{\rm A}$ is turned again set as $C_{\rm {bd}}^{\rm A}=1$,
the connection structure transforms back to that shown in Fig. \ref{fig:snape2}(a).
Thus, these are not kinetically trapped structures but local equilibrium structures.
For high spontaneous curvature at $C_{\rm {bd}}^{\rm A}=1$, the saddle structure at the side of the connection
becomes unstable so that the cylindrical structure directly connects to the type B domain
despite partial exposure of the type B membrane edges.

\section{Summary and Discussion}
\label{sec:sum}

We have found various morphologies of micelles in mixtures of two types of amphiphilic molecules.
For lipid-detergent mixtures, bicelle and worm-like micelles without branches are formed
in dilute solutions.
The detergent-type molecules stabilize the edges of the bilayer disks.
On the other hand, for lipid-lipid mixtures, it is found that 
the worm-like micelles are stably connected
with bicelles.

Two types of connection structures between the bicelle and worm-like micelles are found.
At a medium spontaneous curvature of the type A monolayer,
worm-like micelles of type A molecules are connected with the circular type A domain surrounding the type B domain.
At a high spontaneous curvature, worm-like micelles are directly connected with the type B domain.
Worm-like micelles are  connected with each other though the rim consisting of the type A domain
in the first case (the round-shaped connection) while the absent for the direct connection in the second case.
These octopus-like morphologies resemble the cryo-TEM images of diblock copolymer mixtures \cite{jain04}.
These diblock copolymers have  negligibly small CMC, similar to  phospholipids.
Since the connection points are round in the cryo-TEM images,
the structure is likely the former round-type connection structures;
the rim and octopus arms are made of one type of copolymer with a relatively longer hydrophilic chain.
Our simulation suggests that the octopus-shaped micelles are generally formed
in binary mixtures of negligibly-low-CMC surfactants.
Hence, similar octopus-like micelles can be constructed in binary mixtures of cone-shaped lipids and cylindrical-shaped lipids.

During the solubilization of vesicles in lipid-detergent mixtures,
 octopus-like micelles appear as temporal structures.
Such temporal structures were considered in some experimental studies \cite{walt91,elsa11}.
Our simulation results support the existence of these intermediate structures.

To control micelle morphology, several key quantities require to be tuned.
The spontaneous curvature of the monolayer or the packing parameter\cite{isra11} of the surfactants
determines stable local structure of each phase domain.
The ratio of line tensions of the phase boundary and membrane edges
determines the wetting condition for the membrane edges.
The contact angles of a pore on the phase boundary are given by the Young equation.
At finite CMC,  the excess surfactant can be dissolved from the micelles,
and the composition of surfactants can be adjusted to their stable shapes.
On the other hand, at negligibly low CMC, 
the number of surfactant molecules can be conserved in micelles.
Hence, discoidal and cylindrical structures can coexist in a single micelle, 
where the excess amount of  cone-shaped molecules to surround the circumference form connected worm-like micelles.
The micelle size can be controlled by kinetics as well as 
the volume fractions of the surfactants.
The lipid domain size at the time of vesicle rupture determines  the resulting bicelle size.

In this paper, we focus on micelle formation from a vesicle.
The size and shape of micelles depend on  initial conditions.
Self-assembly from molecules randomly dissolved in solutions
produces smaller sizes of micelles.
Micelle formation by self-assembly or other initial conditions
will be explored in further studies.

\begin{acknowledgments}
This study is partially supported by a Grant-in-Aid for Scientific Research 
on Priority Area ``Molecular Science of Fluctuations toward Biological Functions'' from
the Ministry of Education, Culture, Sports, Science, and Technology of Japan.
\end{acknowledgments}

%\bibliographystyle{apsrev}
%\bibliography{mls,tri}

\end{document}